\documentclass{amsproc}

\usepackage{amsmath, amsthm, amssymb}

\newtheorem{theorem}{Theorem}[section]
\newtheorem{lemma}[theorem]{Lemma}

\theoremstyle{definition}

\newtheorem{heuristic}[theorem]{Heuristic Assumption}

\theoremstyle{remark}

\numberwithin{equation}{section}

\newcommand{\Z}{\mathbb Z}
\newcommand{\F}{\mathbb F}

\begin{document}


\title{Finding primitive elements in finite fields of small characteristic}


%    author one information
\author{Ming-Deh Huang}
\address{Computer Science Department, University of Southern California}
%\curraddr{}
\email{mdhuang@usc.edu}
%\thanks{}

%    author two information
\author{Anand Kumar Narayanan}
\address{Computer Science Department, University of Southern California}
%\curraddr{}
\email{aknaraya@usc.edu}
%\thanks{}

\subjclass[2010]{12E20 (primary),11Y16}
%    The 2010 edition of the Mathematics Subject Classification is
%    now available.  If you are citing a classification from the
%    new scheme, use the following input coding instead.
%\subjclass[2010]{Primary }

\date{}

\begin{abstract}
We describe a deterministic algorithm for finding a generating element of the multiplicative group of the finite field with $p^n$ elements. In time polynomial in $p$ and $n$, the algorithm either outputs an element that is provably a generator or declares that it has failed in finding one. Under a heuristic assumption, we argue that the algorithm does always succeed in finding a generator. The algorithm relies on a relation generation technique in a recent breakthrough by Antoine Joux's for discrete logarithm computation in small characteristic finite fields in $L(1/4,o(1))$ time. For the special case when the order of $p$ in $(\Z/n\Z)^\times$ is small (bounded by $(\log_p(n))^{\mathcal{O}(1)}$), we present a modified algorithm which is reliant on weaker heuristic assumptions.
\end{abstract}

\maketitle

\section{Introduction}
Let $p$ be a prime and $n$ a positive integer. The multiplicative group $\F_{p^n}^{\times}$ of the finite field $\F_{p^n}$ is cyclic and has $\phi(p^n-1)$ generators (also called primitive elements), where $\phi$ is Euler's totient function. Since $\phi(p^n-1) = \Omega(\frac{p^n-1}{\log(\log(p^n-1))})$ \cite{hw}, a large fraction of $\F_{p^n}^{\times}$ are primitive elements. In spite of their abundance, finding one efficiently remains an important open problem. The difficulty partly lies in testing if a given element is a generator and all known algorithms for testing either factor $p^n-1$ or solve an instance of the discrete logarithm problem in $\F_{p^n}^\times$, both of which are believed to be difficult.\\ \\
Even if the question were relaxed and an element of large order is sought, approaches that work in general for every $p$ and $n$ are rare. Gao \cite{gao} presents an algorithm that produces an element of order $\exp(\Omega(\log n)^2/\log(\log (n)))$. Gao's algorithm is efficient conditioned on a conjecture which bears resemblance to our heuristic \ref{heu1}. Voloch \cite{vol} presents an approach suited to small $p$ that finds an element of order $\exp(\Omega(\sqrt{n}))$. Notably, no previous algorithms to compute an element of order exponential in $n$ were known, even if allowed to make heuristic assumptions.\\ \\
There are other constructions that provably find an element of large order, but they only apply to very special $(p,n)$ pairs \cite{vs}\cite{asv}\cite{cheng}\cite{cgw}\cite{chang1}\cite{pop1}\cite{pop2}. For certain $(p,n)$ pairs, von zur Gathen and Shparlinski \cite{vs}  introduced the idea of constructing elements of high order using Gauss periods. Extensions and improvements on their results appear in \cite{asv}\cite{chang1}\cite{pop1}\cite{pop2}. When $n=\frac{p^c-1}{p-1}$ for some $c>1$, Cheng, Gao and Wan \cite{cgw} describe a deterministic algorithm that finds an element of order $\exp(\Omega(\sqrt{p^c}))$ in time polynomial in $p^c$. Voloch \cite{vol2} and Chang \cite{chang2} present constructions based on elements appearing as coordinates of points on certain curves.\\ \\
An alternate relaxation of the question is to find small sets that contain a generator. Davenport \cite{dav} proved that when $p$ is large enough compared to $n$ and $\F_{p^n} = \F_p[\theta]$, the set $\F_p + \theta$ contains a generator of $\F_{p^n}^\times$. Shoup \cite{shoup} extended this result to prove the existence of a subset $A \subseteq \F_{p^n}$ of size polynomial in $p$ and $n$ that contains a generator. Further, the set contains elements of degree bounded by $\mathcal{O}(\log_p(n))$ when represented as polynomials in $\theta$. Shparlinski in \cite{shp1} gave a simpler more efficient construction and in \cite{shp2} further reduced the size of the subset $A$. The question remains on how to identify a generator given a small set that contains one.\\ \\
In recent breakthroughs, Gologlu, Granger, McGuire, Zumbragel \cite{gggz} and Joux \cite{joux} independently devised algorithms that assuming certain widely believed heuristics compute discrete logarithms in small characteristic finite fields faster than previously known. The authors of \cite{gggz} demonstrated their algorithm by computing discrete logarithms in $\F_{2^{1971}}$ which at the time of announcement was a record \cite{gggz1}. Joux's algorithm is the first to compute discrete logarithms in heuristic $L(1/4,o(1))$ time, where $L(\ell,c)$ is defined as $\exp((c+o(1)) (\log(p^n)^\ell)(\log\log(p^n))^{1-\ell})$. All previous algorithms required $L(1/3,o(1))$ time and this speed up allowed Joux \cite{joux2} to compute discrete logarithms in $\F_{2^{4080}}$. Gologlu, Granger, McGuire and Zumbragel \cite{gggz2} then extended the record to $\F_{2^{6120}}$. \\ \\
A remarkable feature shared by the algorithms is that they both consider a small set as the factor base, one that is of size polynomial in the extension degree. Further, if the extensions they consider are obtained by adjoining a root $\zeta$, then the factor base contains the elements that can be represented as linear polynomials in $\zeta$.\\ \\% Assuming their relation generation algorithms succeed, discrete logarithms of the factor base elements can be determined up to a common constant multiple.\\ \\
We propose to use the factor base and relation generation technique in the initial phase of Joux's paper \cite{joux} to efficiently find generators in $\F_{p^n}^\times$. Whereas the algorithm for discrete logarithm computation assumes a given generator of the entire group, our interest is to find such a generator. The relation generation procedure collects multiplicative relations satisfied by the elements in the factor base and is guaranteed to collect enough only under a heuristic assumption. Unlike in discrete logarithm computations, while computing primitive elements it is not straight forward to check if the relations generated suffice and if so to extract from it a primitive element. To this end, we modify both the factor base and the relation generation step and describe how to test if the generated relations suffice and if so to obtain a primitive element. The factor base is chosen such that if the relation generation step is successful, then the collected relations among the elements of the factor base determine a group whose largest invariant factor contains a large cyclic subgroup of $\F_{p^n}^\times$. Further, we can test if the relation generation was successful from the invariant decomposition of the the group determined by the relations and if successful extract a generator of this large cyclic subgroup of $\F_{p^n}$(see section \ref{testing}). Once a generator for this large subgroup is known, a primitive element can be computed. For the aforementioned invariant factor to contain a large cyclic subgroup of $\F_{p^n}^\times$, the factor base does not necessarily have to contain a primitive element. It suffices if the factor base generates the whole multiplicative group, and this is indeed the case as we observe that a result of F.R.K Chung \cite{chung} nicely applies to our situation when the finite field is considered as an extension over a large enough base field.\\ \\
Our algorithm, in time polynomial in $p$ and $n$, either certifiably finds a generator or indicates that it has failed in doing so. Moreover assuming a slightly weaker heuristic assumption than what is implicitly assumed in Joux's method, our algorithm finds a generator in time polynomial in $p$ and $n$ (see Theorem 2.4). In addition to the heuristic reasoning provided in this paper, the success of Joux's method in breaking the record of discrete logarithm computation can be taken as a strong evidence in support of the heuristic assumption.\\ \\
It should be noted that our running time has polynomial dependence on $p$ and not on $\log p$. Thus the algorithm is efficient only in small characteristic.\\ \\  
For instances where $p$ is of small order in $(\Z/n\Z)^\times$, we present a modified algorithm that is simpler to state and reliant on fewer heuristic assumptions.\\ \\
In a recent further advancement \cite{bgjt}, Barbulescu, Gaudry, Joux  and Thome have discovered an algorithm to compute discrete logarithms in $\F_{q^{2n}}^\times$ for $n \leq q$ in $q^{\mathcal{O}(\log n)}$ time based on heuristics. Their result combined with Shoup's \cite{shoup} proof of the existence of small sets containing a primitive element implies a heuristic algorithm to compute primitive elements in $\F_{p^n}$ with quasi-polynomial running time $\left(pn\right)^{\mathcal{O}(\log n)} $. Our algorithm is faster since the running time is polynomial in $p$ and $n$. 
\section{Finding Primitive Elements}
\subsection{Overview of the Algorithm}
The algorithm first proceeds by embedding $\F_{p^n}$ into an extension $\F_{q^{2m}}$ where $q$ is a power of $p$ such that $n \leq q$ and $m$ is a multiple of $n$ such that $q/2 < m \leq q$. In particular, we set $q := p^{\lceil \log_p(n)\rceil}$ and $m$ is chosen as the largest integral multiple of $n$ satisfying $q/2 < m \leq q$. We remark that our choice of embedding field $\F_{q^{2m}}$ is in certain cases larger than the one chosen in Joux's algorithm \cite{joux}.\\ \\
The field $\F_{q^{2m}}$ is constructed as $\F_{q^2}[\zeta]$, where $\zeta$ is a root of an irreducible polynomial $g(x) \in \F_{q^2}[x]$ of degree $m$ that is of a special form. Following Joux, we seek polynomials $h_0, h_1 \in \F_{q^2}[x]$ of low degree such that the factorization of $h(x) := h_1(x)x^q - h_0(x)$ over $\F_{q^2}[x]$ has an irreducible factor of degree $m$ and pick $g(x)$ to be one such irreducible factor of degree $m$. The motivation behind choosing $g$ in this manner is that the identity $h_1(\zeta) \zeta^q - h_0(\zeta)=0$ would later allow us to replace $\zeta^q$ with an expression consisting of the low degree polynomials $h_0(\zeta)$ and $h_1(\zeta)$. For technical reasons explained in section \ref{testing}, we deviate from Joux's algorithm and impose three further restrictions on $h(x)$ (see section \ref{search_section}).\\ \\ 
Once $h_0(x), h_1(x)$ and hence $g(x)$ are chosen, we invoke Joux's relation generation algorithm which picks a small subset of $\F_{q^{2m}}^\times$ as the factor base and finds a set of multiplicative relations satisfied by the elements in the factor base. However, the success of the relation generation algorithm in finding enough relations is reliant on certain heuristic assumptions.\\ \\
We show in section \ref{testing} that if sufficiently many relations are generated, then they yield a primitive element. A theorem of F.R.K Chung assures that the subgroup generated by the factor base contains a primitive element and is an important ingredient in our argument. Further, we devise a sufficient condition on the outcome of the relation generation step that can be tested and that if found true leads to efficient computation of a primitive element $\gamma$ that generates $\F_{q^2}[\zeta]^\times$. \\ \\ 
As a consequence, $\delta := \gamma^{(q^{2m}-1)/(p^n-1)}$ has order $p^n-1$ and generates the multiplicative group of $\F_{p}[\delta] \cong \F_{p^n}$.\\ \\
We assume an explicit representation of $\F_{p^n}$ (see \cite{lenstra}) as an input. That is, a representation of $\F_{p^n}$ as an $\F_p$ vector space with a basis that allows efficient multiplication. For instance, regarding $\F_{p^n}$ as $\F_p[\mu]$ where $\mu$ is a root of a known irreducible degree $n$ polynomial is an explicit representation. Due to Lenstra \cite{lenstra}[Thm 1.2], an isomorphism between two explicit representations of a field of size $p^n$ can be computed deterministically in time polynomial in $n$ and $\log(p)$. Thus a generator for any explicit representation of $\F_{p^n}$ can be found as the image of $\delta$ under an isomorphism.\\ \\
The algorithm is deterministic and it always terminates in time polynomial in $n$ and $p$. We either successfully find a primitive element or declare failure. The algorithm can fail for two reasons, either we fail in finding $g(x)$ of the special form or the relations generated do not suffice. Based on heuristic assumptions, we argue that neither occurs.
\subsection{The Polynomial Search Phase:}\label{search_section}
Let $C$ be a positive integer. We say that an integer is $q^{2C}$-smooth if and only if all its prime factors are at most $q^{2C}$.\\ \\
%\begin{definition}\label{good} 
We define a polynomial $f(x) \in \F_{q^2}[x]$ to be ``good" if and only if the following four conditions are satisfied.
\begin{enumerate}
\item $f(x)$ is square free. 
\item $f(x)$ does not have linear factors.
\item $f(x)$ has an irreducible factor of degree $m$.
\item For every irreducible factor $g^\prime(x)$ of $f(x)$ such that $\deg(g^\prime(x)) \neq m$, $gcd(q^{2\deg(g^\prime)}-1, q^{2m}-1)$ is $q^{2C}$-smooth.
\end{enumerate}
% Consider the factorization of a square free polynomial $f(x) \in \F_{q^2}[x]$, $$f(x) = \prod_{i=0}^{u} f_i$$ into powers of distinct irreducible polynomials $f_i \in \F_{q^2}[x]$ where $\deg(f_i) \geq \deg(f_{i+1})$ for $0 \leq i \leq u$. We say that $f(x)$ is ``good" if and only if $\deg(f_0)= m$ and $\forall 1 \leq i \leq u$, $\deg(f_i(x)) > 1$ and $( q^{2\deg(g_i(x)}-1), q^{2m}-1)$ is $q^{2C}$-smooth. %That is, $f(x)$ is ``good" if and only if in its factorization into powers of irreducible polynomials, there is an irreducible factor of degree $m$ with multiplicity $1$, every other irreducible power has degree at most $m-1$ and there are no linear factors. \\ \\
%\end{definition}
%We say that a polynomial in $\F_{q^2}^\times[x]$ is ``good" if and only if it is square free, has an irreducible factor of degree $m$, and all its other irreducible factors are of degree greater than $1$ and less than $m$.\\ \\
We set a degree bound $D$ and investigate the existence of $h_0,h_1 \in \F_{q^{2}}[x]$ each of degree bounded by $D$ such that $h(x) = h_1(x) x^q-h_0(x)$ is ``good". \\ \\
The existence of ``good" polynomials of the above form requires that $q+D$ is at least $m+2$  for otherwise we are left with a linear factor. To this end, if $m=q$, we assume $D>1$ and if $m=q-1$, we assume $D>0$.\\ \\
For $m>2$ and $r \geq m$, let $N_q(r,m)$ denote the number of polynomials in $\F_{q^2}[x]$ of degree $r \geq m$ that satisfy the first three conditions of being ``good" and let $P_q(r,n)= \frac{N_q(r,n)}{q^{2r}}$ denote the probability that a random polynomial of degree $r$ satisfies the first three conditions of being ``good". Let $s$ and $t$ be non negative integers such that $q+D-m = s(m-1)  + t$, where $t < m-1$. For a positive integer $k$, let $I_k$ denote the number of monic irreducible polynomials in $\F_{q^2}[x]$ of degree $k$.\\ \\
If $t \neq 1$, then $$N_q(q+D) \geq I_m \binom{I_{m-1}}{s} I_t$$
since we can chose an irreducible polynomial of degree $m$, $s$ irreducible polynomials of degree $m-1$ and one irreducible polynomial of degree $t$ and take their product to get a polynomial of degree $q+D$.
By substituting the lower bound $$I_k \geq \frac{q^k}{k} - \frac{q(q^{k/2}-1)}{(q-1)k}$$ in the above expression we get
%$$N_q(q+d,m) \geq \left( \frac{q^m}{m} - \frac{q(q^{m/2}-1)}{(q-1)m}\right)\left( \binom{\frac{q^{(m-1)}}{m-1} - \frac{q(q^{(m-1)/2}-1)}{(q-1)(m-1)}}{s}\right)\left( \frac{q^t}{t} - \frac{q(q^{t/2}-1)}{(q-1)t}\right)$$
$$P_q(q+D,m) = \frac{N_q(q+D,m)}{q^{2(q+D)}} \geq \frac{1}{m (m-1)^s t s! } \left( 1 - \mathcal{O}\left(\frac{1}{q^{t}}\right)\right).$$\\
Likewise, when $t=1$, it follows that $s \geq 1$ and we obtain $$N_q(q+D,m) \geq I_m \binom{I_{m-1}}{s-1} I_{m-2} I_{t+1}$$ $$\Rightarrow P_q(q+D,m) \geq \frac{1}{m (m-1)^{s-1} (m-2) (t+1) (s-1)! } \left( 1 - \mathcal{O}\left(\frac{1}{q^{t+1}}\right)\right).$$\\ \\
If we were to assume that a random polynomial of the form $h_1(x)x^q - h_0(x)$, where $h_0$ and $h_1$ are of degree at most $D$ satisfies the first three conditions of being ``good" with probability $P_q(q+D,n)$, then since $s = \mathcal{O}(D/m)$ choosing $$D = \Theta(\log_{q^2}(m (m-1)^s t s!)) = \Theta(1)$$ is sufficient to ensure the existence of $h_0$ and $h_1$ such that $h(x)$ is square free, has a degree $m$ factor and no linear factors.\\ \\
Heuristically it is likely that a large fraction of polynomials that satisfy the first three constraints also satisfy the fourth constraint on being ``good". \\ \\
For a polynomial that satisfies the first three conditions, if each of its factors excluding its degree $m$ factor is either of degree prime to $m$ or of degree bounded by $C$, then it is likely to satisfy the fourth condition.\\ \\
Consider positive integers $m^\prime, s^\prime$ and $t^\prime$ such that $m^\prime > m/2$, $t^\prime > 1$,  $q+D-m = s^\prime m^\prime + t^\prime$, $\gcd(m^\prime,m)=1$ and either $\gcd(t^\prime,m)=1$ or $t < C$. For such a choice, $\gcd(q^{2m^\prime}-1,q^{2m}-1)$ and $\gcd(q^{2t^\prime}-1,q^{2m}-1)$ are both likely to be $q^{\mathcal{O}(1)}$-smooth. Hence by taking an irreducible polynomial of degree $m$, $s^\prime$ irreducible polynomials of degree $m^\prime$ and an irreducible polynomial of degree $t^\prime$, we can construct a ``good" polynomial. From an analysis similar to the above computation of $P_q(q+D,m)$, we can conclude heuristically that choosing $D = \Theta(1)$ and $C =\Theta(1)$ are sufficient to guarantee the existence of the ``good" polynomials that we seek.
\begin{heuristic}\label{heu1} There exists positive integers $D,C$ such that for all prime powers $q$ and for all positive integers $2 < m \leq q$, there exists $h_0,h_1 \in  \F_{q^2}[x]$ of degree bounded by $D$ such that $h_1(x)x^q-h_0(x)$ is square free, has an irreducible factor (call $g(x)$) of degree $m$, and for each irreducible factor $g^\prime(x)$ of $h(x)/g(x)$,  $\deg(g^\prime) > 1$ and $\gcd(q^{2\deg(g^\prime)}-1, q^{2m}-1)$ is $q^{2C}$-smooth.
\end{heuristic}
\textbf{Search for $h_0(x), h_1(x)$ and $g(x)$: }\textit{Fix constants $C,D$. Enumerate candidates for $h_0, h_1\in \F_{q^2}[x]$ with each of their degrees bounded by $D$. For each candidate pair $(h_0 , h_1)$, factor $h(x) = h_1(x)x^q-h_0(x)$. If $h(x)$ is ``good", output $h_0, h_1$ and the factor of degree $m$ and stop. If no such candidates are found, declare failure.}\\ \\
The search algorithm terminates after considering at most $q^{2(D+1)} = q^{\mathcal{O}(1)}$ candidate pairs. Factoring each candidate $h_1(x)x^q-h_0(x)$ takes time polynomial in the degree $q+D$ and $p$ using Berlekamp's deterministic polynomial factorization algorithm\cite{ber}. All four conditions of being good can be checked efficiently given the degrees of the irreducible factors in the factorization of $h$. Thus, the search for $h_0, h_1$ and hence $g$ of the desired takes at most $q^{\mathcal{O}(1)}$ time.
\subsection{Small Generating Set} \label{gen}
We next choose a small subset $S \subseteq \F_{q^2}[\zeta]$ that generates $\F_{q^{2}}[\zeta]^\times$. F.R.K Chung proved that for all prime powers $s$, for all positive integers $r$ such that $(r - 1)^2<s$, for all $\mu$ such that $\F_{s^r} = \F_{s}[\mu]$, the set $\F_s + \mu$ generates $\F_{s^r}^\times$ \cite[Thm. 8]{chung}\cite[Ques 1.1]{wan}. Since $m \leq q$, setting $S :=  \F_{q^2} + \zeta $ ensures that the subgroup generated by $S$, $\langle S \rangle = \F_{q^{2}}[\zeta]^{\times}$.\\ \\
Given that $\langle S \rangle = \F_{q^{2m}}^\times$, the next step is to determine the relations satisfied by the elements in $S$ so that we can determine $\F_{q^{2}}[\zeta]$ as the free abelian group generated by $S$ modulo the relations.\\ \\
For a technical reason, $S$ is first extended to the set $F := h_1(\zeta) \cup \{\lambda\} \cup S$, where $\langle \lambda \rangle = \F_{q^2}^\times$ . An identity in $\F_{q^{2m}}^\times$ of the form $\prod_{\beta \in F} \beta^{e_\beta} = 1$ for integers $e_\beta$ is called as a relation and it can be identified with the relation vector $(e_\beta, \beta \in F)$ indexed by elements in $F$.
\subsection{Joux's Relation Generation Algorithm}\label{relation}
The relation search step begins with the following identity over $\F_{q^2}[x]$ $$\prod_{\alpha \in \F_{q}}{x-\alpha} = x^q - x.$$
For $(a,b,c,d) \in \F_{q^2}^{4}$ such that $ad-bc \neq 0$, the substitution $x \mapsto \frac{a\zeta+b}{c\zeta+d}$ yields $$ \prod_{\alpha \in \F_q} \frac{(a-\alpha c)\zeta + (b-\alpha d)}{(c \zeta +d)^{q}} = \frac{(c \zeta + d)(a \zeta + b)^q - (a\zeta+b)(c\zeta+d)^q }{(c\zeta+d)^{q+1}}$$
$$\Rightarrow (c\zeta+d) \prod_{\alpha \in \F_q} ((a-\alpha c)\zeta + (b-\alpha d)) = (c \zeta + d)(a \zeta + b)^q - (a\zeta+b)(c\zeta+d)^q. $$
Linearity of raising to the $q^{th}$ power implies $$ (c\zeta+d) \prod_{\alpha \in \F_q} ((a-\alpha c)\zeta + (b-\alpha d)) = (c \zeta + d)(a^q \zeta^q + b^q) - (a \zeta+b)(c^q \zeta^q+d^q).$$
By substituting $\zeta^q = \frac{h_0(\zeta)}{h_1(\zeta)}$, the right hand side becomes $$\frac{(ca^q-ac^q) \zeta h_0(\zeta) + (da^q-bc^q) h_0(\zeta) + (cb^q - ad^q) \zeta h_1(\zeta) + (db^q-bd^q) h_1(\zeta)}{h_1(\zeta)}.$$
Consider the numerator of the above expression as the polynomial $$N(x) : =  (ca^q-ac^q) x h_0(x) + (da^q-bc^q) h_0(x) + (cb^q - ad^q) x h_1(x) + (db^q-bd^q) h_1(x) $$ evaluated at $\zeta$. The degree of $N(x)$ is bounded by $d+1$. If $N(x)$ factors in to linear factors over $\F_{q^2}[x]$, then we get the following relation in $\langle F \rangle$ $$(c\zeta+d) h_1(\zeta) \prod_{\alpha \in \F_q} ((a-\alpha c)\zeta + (b-\alpha d))  = n(\zeta).$$
The above expression can be written as a product of an element $\mu \in \F_{q^2}^\times$ times $h_1(\zeta)$ times a fraction of products of monic linear polynomials in $\zeta$ over $\F_{q^2}$ being equal to $1$. By expressing the element $\mu$ in $\F_{q^2}^\times$ as a power of $\lambda$ by computing a discrete logarithm over $\F_{q^2}^\times$, we indeed get a relation in $\langle F \rangle$.\\ \\
The reason for choosing to work over $\F_{q^2}$ instead of $\F_q$ is that for every choice of $a,b,c,d \in \F_q$, the relation it yields becomes $\zeta^q-\zeta = \prod_{\alpha \in \F_q} (\zeta-\alpha)$. Thus, we have to work over an extension of $\F_q$ where the $q^{th}$ power map would be non trivial and $\F_{q^2}$ is the smallest such extension.\\ \\
\textbf{Relation Generation:} \textit{For every $(a,b,c,d) \in \F_{q^2}^{4}$ such that $ad-bc \neq 0$, compute the numerator $N(x)$ and if it factors into linear factors over $\F_{q^2}[x]$, add the relation as a row to the relation matrix $R$.\\ \\
Add the relation corresponding to the identity $\lambda^{q^2-1} = 1$ to $R$.}\\ \\
%Add the relations corresponding to the identities $\forall u \in F -\{\lambda\}, u^{q^{2m}-1} = 1$ and $\lambda^{q^2-1} = 1$ to the relation matrix $R$.}\\ \\
The relation generation step can be performed in $q^{\mathcal{O}(1)}$ time since the number of choices for $(a,b,c,d)$ is at most $q^{\mathcal{O}(1)}$ and factoring the numerator polynomial using Berlekamp's deterministic factoring algorithm takes $q^{\mathcal{O}(1)}$ time as the numerator polynomial is of constant degree. We have to express the constant $\F_{q^2}^\times$ factor in the relation as a power of $\lambda$, but that can be accomplished by solving the discrete logarithm in $\F_{q^2}^\times$ exhaustively in $\mathcal{O}(q^2)$ time. 
\subsection{Testing}\label{testing}
Let $R$ be the $N$ by $|F|$ matrix consisting of the relation vectors as rows and $\Gamma_R$ the $\Z$-lattice generated by the rows of $R$. The Smith normal form of $R$ gives the decomposition of $\Z^{|F|}/\Gamma_R$ into its invariant factors $$\Z^{|F|}/\Gamma_R = \langle e(1) \rangle \oplus \langle e(2) \rangle \oplus \ldots \oplus \langle e(|F|) \rangle  \cong  \Z/d_1\Z \oplus \Z/d_2\Z \oplus \ldots \oplus \Z/d_{|F|}\Z $$
where for $1\leq i \leq |F|$, $e(i) \in \Z^{|F|}$ denotes a relation vector and $d_i$ the order of $e(i)$ in $\Z^{|F|}/\Gamma_R$ and for $1 \leq i <|F|$, $d_i \mid d_{i+1}$.\\ \\
For a polynomial $f(x) \in \F_{q^2}[x]$, let $\F_f$ denote the ring $\F_{q^2}[x]/\left( f(x)\F_{q^2}[x]\right)$.\\ \\% and let $\F_f^\times/\ell: = \F_f^\times/(\F_f^\times)^\ell$. 
Let $h= \prod_{i=0}^k g_i(x)$ be a factorization of $h(x)$ into distinct irreducible polynomials in $\F_{q^2}[x]$. Without loss of generality, let $g_0(x)=g(x)$.\\ \\
While our objective in the relation generation step was to collect relations in $\F_g^\times$, the relations generated are in fact satisfied in $\F_{g_i}^\times$ for every $0 \leq i \leq k$. It is to break this symmetry and focus on $\F_{g}^\times$ that we insist that $\forall 1 \leq i \leq k$, $\gcd(q^{2\deg(g_i)}-1, q^{2m}-1)$ is $q^{2C}$-smooth.\\ \\
The fact that the relations generated hold in $\F_{g_i}^\times$ for every $0 \leq i \leq k$ is also of concern to Joux's algorithm for computing discrete logarithms. This was also observed independently by \cite{cwz}.\\ \\
For a non constant polynomial $f(x) \in \F_{q^2}[x]$ dividing $h(x)$, let $\Gamma_f$ denote the relation lattice of the subgroup of $\F_{f}^\times$ corresponding to the generating set $$F_f: =  \{\mu\} \cup \{h_1(x) \mod f(x)\} \cup \{x + \theta \mod f(x) , \theta \in \F_{q^2}\}.$$
That is, $$\Gamma_f = \left\{ (z_\beta)_{\beta \in F_f}  \in \Z^{|F|} | \prod_{\beta \in F_f}\beta^{z_\beta} =1 \right\}.$$
The relation lattice generated $\Gamma_R$ is contained in $\Gamma_h$ which is in turn contained in $\Gamma_g$ and we have the natural surjection $$ \Z^{|F|}/\Gamma_R \twoheadrightarrow \Z^{|F|} /\Gamma_g.$$
Recall F.R.K Chung's theorem that for all prime powers $s$, for all positive integers $r$ such that $(r - 1)^2<s$, for all $\mu$ such that $\F_{s^r} = \F_{s}[\mu]$, the set $\F_s + \mu$ generates $\F_{s^r}^\times$ \cite[Thm. 8]{chung}\cite[Ques 1.1]{wan}. Since $\deg(g(x)) \leq q$, F.R.K Chung's theorem implies that $\Z^{|F|}/\Gamma_g \cong \F_g^\times$\\ \\
Thus, the natural reduction map $\varphi : \Z^{|F|}/\Gamma_R \twoheadrightarrow \F_g^\times$ is surjective. For $1 \leq i <|F|$, let $\pi_{i}$ denote $\varphi(e(i)) = \prod_{\beta \in F} \beta^{e(i)_\beta}$.\\ \\
If $h$ were to have a linear factor, then the relation generation step will not relate that linear factor to the rest of the linear polynomials in the factor base. As a result, we would have to exclude that linear factor from the factor base and F.R.K Chung's theorem would no longer apply. It is to circumvent this that we insisted that $h$ have no linear factors.\\ \\
We next prove a lemma which states a condition on $\Z^{|F|}/\Gamma_R$ that guarantees that our relation generation step has collected enough enough relations to extract an element of large order in $\F_g^\times$. From this large order element we will eventually compute a primitive element.   
\begin{lemma}\label{sufficient} If $\gcd(d_{|F|-1},q^{2m}-1)$ is $q^{2C}$-smooth, then there exists a $q^{2C}$-smooth number $B$ such that the order of  $\varphi(e(|F|))$ in $\F_g^\times$ is divisible by $\frac{q^{2m}-1}{B}$.
\end{lemma}
Assume $\gcd(d_{|F|-1},q^{2m}-1)$ is $q^{2C}$-smooth. From the Smith normal form, we have the invariant factor decomposition $$\Z^{|F|}/\Gamma_R = \bigoplus_{j=1}^{|F|} \langle e(j) \rangle$$ where $d_j$ is the order of $e(j)$ in $\Z^{|F|}/\Gamma_R$.\\ \\% and for $1 \leq i <|F|$, $d_i \mid d_{i+1}$.\\ \\
Since $\left|\varphi\left(\bigoplus_{j=1}^{|F|-1} \langle e(j) \rangle\right) \right| = \prod_{j=1}^{|F|-1} \left|\varphi\left(\langle e(j) \rangle\right)\right|$ divides $\prod_{j=1}^{|F|-1} d_j$ and $d_j \mid d_{j+1}$ for $1 \leq j <|F|-1$, it follows that $\gcd\left(\left|\varphi\left(\bigoplus_{j=1}^{|F|-1} \langle e(j) \rangle\right) \right|,q^{2m}-1\right)$ is $q^{2C}$-smooth. \\ \\
Since $\varphi(\Z^{|F|}/\Gamma_R) = \F_g^\times$ and $\F_g^\times$ is cyclic of order $q^{2m}-1$, there exists a $q^{2C}$-smooth number $B$ such that the order of  $\varphi(e(|F|))$ in $\F_g^\times$ is divisible by $\frac{q^{2m}-1}{B}$. $\square$\\ \\
We next show if the relation generation is successful in computing the relation lattice of $\Gamma_h$ in its entirety, then the condition stated in lemma \ref{sufficient} is satisfied.
\begin{lemma}\label{lattice} If $\Gamma_R = \Gamma_h,$ then $\gcd\left(d_{|F|-1},q^{2m}-1\right)$ is $q^{2C}$-smooth.
\end{lemma}
Let $v$ denote the largest factor of $q^{2m}-1$ that is $q^{2C}$-smooth and let $L = (q^{2m}-1)/v$.
Since $h$ is square free, $$\F_h^\times \cong \F_g^\times \times \prod_{i=1}^k \F_{g_i}^\times.$$
Let $\langle F_h \rangle$ denote the subgroup of $\F_h^\times$ generated by $F_h$. We have the inclusion $$\psi: \langle F_h \rangle \hookrightarrow \F_g^\times \times \prod_{i=1}^k \F_{g_i}^\times$$ $$\ \ \ \ \  \alpha \longmapsto \alpha_g \prod_{i} \alpha_{g_i} $$
Since the projection from $\langle F_h \rangle$ to $\F_g^\times$ is surjective, there exists a $\beta \in \langle F_h \rangle$ whose projection $\beta_g$ in $\F_g^\times$ is of order $q^{2m}-1$.\\ \\
%The order of $\beta \in \langle F_h\rangle$ equals the least common multiple of $\{ \beta_g\} \cup \{\beta_{g_i}| 1 \leq i \leq k\}$.
The order of $\beta \in \langle F_h\rangle$ is divisible by the order of its projection $\beta_g \in \F_g^\times$. Hence $\langle F_h\rangle$ has an element of order $q^{2m}-1$ which implies that we have an inclusion $$\Z/L\Z \hookrightarrow \langle F_h \rangle$$ and hence $L$ divides $\left| \langle F_h\rangle\right|$.\\ \\
Since $\langle F_h \rangle \hookrightarrow \F_g^\times \times \prod_{i=1}^k \F_{g_i}^\times$, $\left| \langle F_h\rangle\right|$ divides $(q^{2m}-1)\prod_{i=1}^k (q^{2\deg(g_i)}-1)$.\\ \\
Since $gcd(q^{2\deg(g_i)}-1,q^{2m}-1)$ is $q^{2C}$-smooth for $g_i \neq g$, it follows that there exists integers $w,y$ such that $w$ is $q^{2C}$-smooth, $\gcd(L,y) = 1$ and $\left| \langle F_h\rangle\right| = L w y$.\\ \\
For every prime $\ell$ dividing $L$, the $\ell$-primary component of $\langle F_h\rangle$ is cyclic since $\Z/L\Z \hookrightarrow \langle F_h \rangle$ and $\left| \langle F_h\rangle\right|$ is $L$ times a factor relatively prime to $L$. Hence in the Smith normal form of $\langle F_h \rangle$, for every prime $\ell$ dividing $L$, the $\ell$-primary component of $\langle F_h\rangle$ is contained in the largest invariant factor. In particular, the largest invariant factor has order divisible by $L$.\\ \\
Since $\left| \langle F_h \rangle\right| = L w y$, it follows that the second largest invariant factor of $\langle F_h \rangle$ has order dividing $wy$. Since $w$ is $q^{2C}$-smooth and $\gcd(L,y)=1$, $\gcd(wy,q^{2m}-1)$ is $q^{2C}$-smooth.\\ \\
If $\Gamma_R = \Gamma_h$, then $\Z^{|F|}/\Gamma_R \cong \langle F_h \rangle$ and the order $d_{|F|-1}$ of the second largest invariant factor of $\Z^{|F|}/\Gamma_R$ divides $wy$. Thus $\gcd(d_{|F|-1},q^{2m}-1)$ is $q^{2C}$-smooth. $\square$.\\ \\
\textbf{Testing Phase:} \textit{Compute the Smith normal form of $R$ and if $\gcd(d_{|F|-1},q^{2m}-1)$ is  $q^{2C}$-smooth, output $\pi_{|F|}$. Else, declare failure.}\\ \\
The Smith normal form computation can be performed in $q^{\mathcal{O}(1)}$ time since $R$ has at most $\Theta(q^3)$ rows, at most $q^2+2$ columns and each entry is an integer bounded by $q^2$.\\ \\
If the testing phase is successful, we can extract a primitive element of $\F_g^\times$ from the output $\pi_{|F|}$ of the testing phase as follows. Recall that $v$ is the largest $q^{2C}$-smooth factor of $q^{2m}-1$. If $\mu \in \F_g^\times$ is of order divisible by $v$, then $\mu\pi_{|F|}$ is a primitive element in $\F_g^\times$.\\ \\
Shoup \cite{shoup} proved that there exists a constant $C_1$ such that $P:=\{f(\zeta)| f\in \F_q^2[x], \deg(f) \leq C_1 \log_q(m)\}$ contains a generator of $\F_{g}^\times$. In particular, $P$ has an element of order divisible by $v$.\\ \\
Since $C$ is a constant, $v$ can be computed in time polynomial in $q$. For an $\epsilon \in P$, we can check if it has order divisible by $v$ by verifying that $\epsilon^{(q^{2m}-1)/v} \neq 1$. By exhaustively searching, we can find an element $\mu \in P$ of order divisible by $v$ in time polynomial in $\left|P\right|$ which is polynomial in $q$.
\subsection{Relation Generation Heuristic}
In this subsection, we argue under a heuristic assumption that the relation generation algorithm does indeed produce enough relations to successfully extract a primitive element.\\ \\
We begin by counting the number of relations that we could obtain by counting the possible choices for $(a,b,c,d)$ in the relation generation algorithm.\\ \\
For an $e \in \F_{q^2}^\times$, the substitutions $x \mapsto \frac{a \zeta + b}{c \zeta + d}$ and $x \mapsto \frac{ae \zeta + be}{ce \zeta + de}$ are identical and will lead to the same relation. Thus, the possible choices for $a,b,c,d \in \F_{q^2}$, that could lead to distinct relations can at best be identified with elements in $PGL(2,q^2)$.\\ \\ 
Further, the relations corresponding to an element in $PGL(2,q^2)$ and its product with an element in $PGL(2,q)$ are off by the relation corresponding to the identity $\zeta^q-\zeta = \prod_{\alpha \in \F_q} (\zeta-\alpha)$. \\ \\ 
Thus the number of possible choices for $a,b,c,d$ can be identified with elements in the group $PGL(2,q^2)/PGL(2,q)$ which has cardinality $q(q^2+1) = \Theta(q^3)$.\footnote{We would like to thank Antoine Joux for pointing out the need to mod out by $PGL(2,q)$.}\\ \\
The probability that a random polynomial of degree at most $D+1$ factors into linear factors is roughly $\frac{1}{(D+1)!}$ \cite{pgf}. If the numerator polynomials $N(x)$ that appear in the relation generation phase behave as random polynomials of the same degree with respect to their probability of splitting in to linear polynomials, then the expected number of trials required to get a relation is $(D+1)!$.  Since $D$ is a constant independent of $q$ and $n$, the expected number of rows of $R$ is a constant fraction of $\Theta(q^3)$.\\ \\
Since the dimension of the lattice $|F|$ is at most  $q^2+2$ and $\Gamma_R$ is the lattice generated by $\Theta(q^3)$ points, it is overwhelmingly likely that $\Gamma_R = \Gamma_h$, which makes the weaker claim of the heuristic \ref{relation_lattice} below even more plausible.
\begin{heuristic}\label{relation_lattice}
The generated relation lattice $\Gamma_R$ is large enough to ensure that the greatest common divisor of $q^{2m}-1$ and the cardinality of the second largest invariant factor of $Z^{|F|}/\Gamma_R$ is $q^{2C}$-smooth.
\end{heuristic}
To summarize, our algorithm either certifiably finds a generator or indicates that it has failed in doing so. If the heuristics \ref{heu1} and \ref{relation_lattice} are true, then the algorithm finds a generator in time polynomial in $q$ which is a polynomial in $p$ and $n$. 
\subsection{Reducing the Problem of Finding Primitive Elements to a Conjecture}
Since the generated relation lattice $\Gamma_R$ depends on the choice of the polynomials $h_0$, $h_1$ and $g$, heuristic \ref{relation_lattice} implicitly claims that for every choice of $h_0$, $h_1$ and $g$, the relation generation step succeeds in determining a primitive element. This assumption can be weakened significantly by using the following modified testing phase.\\ \\
\textbf{Modified Testing:} \textit{Compute the Smith normal form of $R$ and if $\gcd(d_{|F|-1},q^{2m}-1)$ is  $q^{2C}$-smooth, output $\pi_{|F|}$. Else, continue with the search for a new choice of $h_0$ and $h_1$.}\\ \\
The modified testing phase implies the following theorem.
\begin{theorem} If there exists positive integers $D,C$ such that for all prime powers $q$ and for all positive integers $q/2 < m \leq q$, there exists $h_0,h_1 \in  \F_{q^2}[x]$ of degree bounded by $D$ such that $h(x) = h_1(x)x^q-h_0(x)$ is square free, has an irreducible factor (call $g(x)$) of degree $m$, and for each irreducible factor $g^\prime(x)$ of $h(x)/g(x)$,  $\deg(g^\prime) > 1$ and $\gcd(q^{2\deg(g^\prime)}-1, q^{2m}-1)$ is $q^{2C}$-smooth and the generated relation lattice $\Gamma_R$ corresponding to $h_0,h_1$ is large enough to ensure that the greatest common divisor of $q^{2m}-1$ and the cardinality of the second largest invariant factor of $Z^{|F|}/\Gamma_R$ is $q^{2C}$-smooth, then a generator for $\F_{p^n}$ can be found deterministically in time polynomial in $p$ and $n$.
\end{theorem}
\subsection{The special case when $p$ is of small order in $(\Z/n\Z)^\times$.}
For the special case when $ord_n(p)$, the order of $p$ modulo $n$ is $(\log_p n)^{\mathcal{O}(1)}$, we present a modification to the algorithm that results in a procedure that has a greater guarantee of success while assuming less.\\ \\
In the initial step, set $q:=p^{ord_n(p)}$ and embed $\F_{p^n}$ in to $\F_{q^{2(q-1)}}$.\\ \\
We skip the search phase and instead set $h_1(x) = 1$ and $h_0(x) = \lambda x$ where $\langle \lambda \rangle = \F_{q^2}\times$. Such an $\lambda$ can be found in $\mathcal{O}(q)$ time by exhaustive searching. Since $h(x) = h_1(x) x^q- h_0(x)=x(x^{q-1}-\lambda)$, where $(x^{q-1}- \lambda)$ is irreducible of degree $q-1$, set $g(x) = x^{q-1}-\lambda$. \\ \\
This choice of $h(x)$ violates the requirements of the search phase of our algorithm since it has a linear factor $x$. The concern is that as a consequence we have to leave out $x \mod g(x)$ from the factor base. However, adding the relation $x^{q-1} \lambda^{-1} = 1 \mod g(x)$ to our relation generation step allows the inclusion of $x \mod g(x)$ in our factor base $F$ and the correctness of the algorithm is not affected.\\ \\
Since the degrees of $h_1$ and $h_0$ are at most $1$, the numerator $N(x)$ that appears in the relation search is of degree at most $2$.\\ \\
If the numerators $N(x)$ behave as random polynomials of degree $2$ in terms of factorization, then they factor with probability $\frac{1}{2}$. Thus, we expect to get at least $q(q^2+1)/2$ relations. In fact, we can prove that we get at least $2q^2+2q-1$ relations.\\ \\
Consider the upper triangular subgroup $G_U$ of $PGL(2,q^2)/PGL(2,q)$, that is, the subgroup whose elements have a representative of the form
\[ \left( \begin{array}{cc}
a & b  \\
0 & 1   \end{array} \right)\] where $a \in \F_{q^2}^\times, b \in \F_{q^2}$. The cardinality of $G_U$ is $((q^2-1)q^2)/((q-1)q) = q^2+q$.\\ \\
For an element in $G_U$ corresponding to an $a \in \F_{q^2}^\times$ and a $b\in \F_{q^2}$, the numerator polynomial $n(x)$ we obtain is the linear polynomial $$(a^q \eta - a)x + (b^q-b).$$
Thus, we are guaranteed at least $q^2+q$ relations.\\ \\
Likewise, by considering the subgroup $G_L $ of $PGL(2,q^2)/PGL(2,q)$ consisting of elements with a lower triangular  representative, we get $q^2+q-1$ more relations.\\ \\
Thus far we have made no heuristic assumptions for this special case. The only assumption we make is that $\Z^{|F|}/\Gamma_R$ is large enough to ensure that the testing phase is successful. The dimension of the relation lattice $\Gamma_h$ is $q^2+1$ and we get at least $2q^2+2q-1$ distinct relations. If the relations that we obtain are modeled as being drawn independently at random from $\Gamma_h$, then with overwhelming probability $\Gamma_R = \Gamma_h$.\\ \\
As a final remark, instead of restricting the factor base $F$ to monic linear polynomials in $\delta$, we could also include the evaluations of quadratic irreducible polynomials in $\F_{q^2}[x]$ at $\delta$, but only those that appear as factors of the $N(x)$ during the relation search. Further, the first time a degree two element is encountered, it can be expressed in terms of a product of linear factors. If a quadratic factor reappears then it implies a new relation between products of linear factors. 
\section{Acknowledgements}
We would like to thank Antoine Joux and Igor Shparlinski for their comments and suggestions on an earlier version of this paper.

\bibliographystyle{amsplain}

\end{document}